\documentclass[manuscript]{acmart}

\AtBeginDocument{%
  \providecommand\BibTeX{{%
    \normalfont B\kern-0.5em{\scshape i\kern-0.25em b}\kern-0.8em\TeX}}}

\setcopyright{acmcopyright}
\copyrightyear{2022}
\acmYear{2022}
\acmDOI{XXXXXXX.XXXXXXX}

\acmConference[HCDS@CHI'22]{Make sure to enter the correct
  conference title from your rights confirmation email}{April 23,
  2022}{Online}
\acmPrice{15.00}
\acmISBN{978-1-4503-XXXX-X/18/06}
\usepackage{tcolorbox}
\usepackage{xcolor}
\usepackage{wrapfig}

\begin{document}

\title{Rankers, Rankees, \& Rankings: Peeking into the Pandora's Box from a Socio-Technical Perspective}

% \affiliation{%

\author{Jun Yuan}
\affiliation{%
  \institution{New Jersey Institute of Technology}
  \country{USA}}
\email{jy448@njit.edu}

\author{Julia Stoyanovich}
\affiliation{%
  \institution{New York University}
  \country{USA}}
\email{stoyanovich@nyu.edu}

\author{Aritra Dasgupta}
\affiliation{%
  \institution{New Jersey Institute of Technology}
  \country{USA}}
\email{aritra.dasgupta@njit.edu}

% \affiliation{%

% \affiliation{%

% \affiliation{%

% \affiliation{%

% \affiliation{%

% \affiliation{%

\renewcommand{\shortauthors}{Yuan, Stoyanovich, and Dasgupta.}


\begin{abstract}
Algorithmic rankers have a profound impact on our increasingly data-driven society. From leisurely activities like the movies that we watch, the restaurants that we patronize; to highly consequential decisions, like making educational and occupational choices or getting hired by companies-- these are all driven by sophisticated yet mostly inaccessible rankers. A small change to how these algorithms process the rankees~(i.e., the data items that are ranked) can have profound consequences. For example, a change in rankings can lead to deterioration of the prestige of a university or have drastic consequences on a job candidate who missed out being in the list of the preferred top-$k$ for an organization. This paper is a call to action to the human-centered data science research community to develop principled methods, measures, and metrics for studying the interactions among the socio-technical context of use, technological innovations, and the resulting consequences of algorithmic rankings on multiple stakeholders. Given the spate of new legislations on algorithmic accountability, it is imperative that researchers from social science, human-computer interaction, and data science work in unison for demystifying \textit{how} rankings are produced, \textit{who} has agency to change them, and \textit{what} metrics of socio-technical impact one must use for informing the context of use.
\end{abstract}

\maketitle

\section*{Position Context: Socio-Technical Impact of Rankings}

% \aritra{when you revise the Intro, try to follow the vocabulary of rankers, rankings, and rankees. Remember, there are also decision-makers as stakeholders, who could be different than rankees: hiring managers, for example. We also need to highlight cases were people are ranked and the high consequence settings of rankings on their profession and life choices}
Algorithmic rankers that learn from data are now ubiquitous and they influence a myriad socio-technical contexts of use.
% A search engine is a ranker that updates every millisecond to rank web pages according to keywords.
For example, Artificial Intelligence~(AI)-based hiring systems employ rankers that match applicants with relevant job profiles and help companies assess quality of fit.
Yet, to the job candidate, who is one in a teeming million of data items which are ranked~(we term them \textit{rankees}), the rankers are obscure and inaccessible. 
\begin{figure*}
    \begin{center}
    \vspace{-10pt}
    \includegraphics[width=\textwidth]{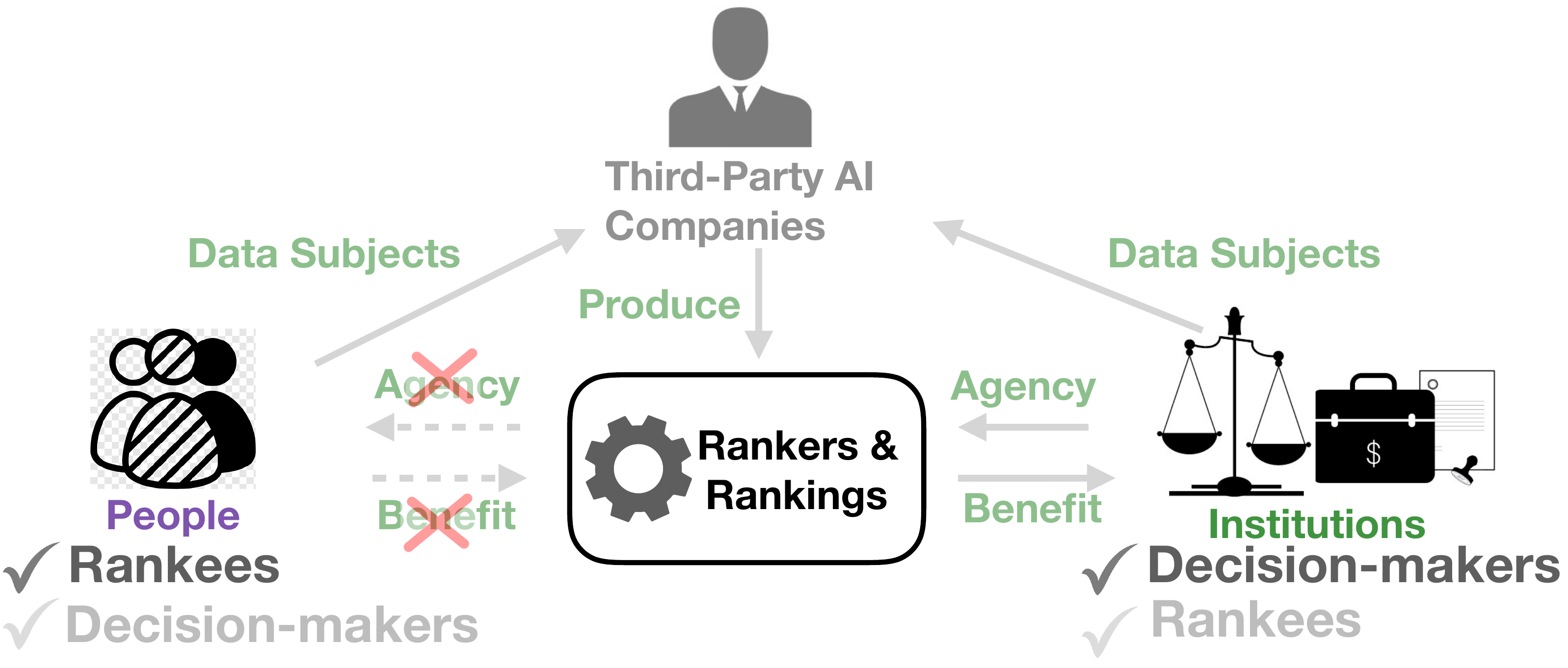}
    \end{center}
    \vspace{-0.1cm}
    \caption{\small{The ecosystem of ranking producers and consumers.}}
    \vspace{-0.3cm}
    \label{fig:rank}
\end{figure*}

% Some rankers have a simpler form, like a (typically linear) weighted summation function for a collection of data items described by multiple attributes, while others are learned.
As another example, US News~\cite{USnews} collects data from educational institutions about their own performance and about their ratings of peers, in order to produce institutional rankings. Here, high rankings are indicative of societal prestige, where elite universities at the top are in demand, often more expensive, and highly selective. People, including students and parents, are consumers of these rankings but have little agency %or stake 
in how they are constructed. These scenarios, indicative of socio-technical inequity, are summarized in Figure~\ref{fig:rank}; they serve as basis for our position statements in this paper, which are also informed by the team's prior work and experience with design and applications of responsible algorithmic rankers~\cite{DBLP:journals/pvldb/AsudehJMS18,DBLP:conf/sigmod/AsudehJS019,DBLP:journals/debu/AsudehJS19,DBLP:conf/sigmod/GuanAMJSM019,DBLP:conf/edbt/StoyanovichYJ18,DBLP:conf/cikm/SunAJHS19,DBLP:conf/ijcai/YangGS19,DBLP:conf/ssdbm/YangS17,DBLP:conf/sigmod/YangSAHJM18,DBLP:conf/forc/YangLS21,DBLP:conf/webdb/StoyanovichJG15,DBLP:journals/corr/abs-2103-14000}.

\begin{tcolorbox}
%As part of proposed work, we will answer the following questions:
\textbf{Statement 1:} Socio-technical practices of ranking-based decision-making need to be informed by a combination of qualitative field studies and quantitative analysis of the risks and benefits of algorithmic rankers, in particular, with respect to rankees who lack agency and visibility into how their data is used for decision-making.
\end{tcolorbox}

% \aritra{the section name should abstract a particular position statement we make. This is just a sample}

% This section defines the different stakeholders, their involvements to a ranking process, and potential issues.
% From the process of rankers, rank rankees to produce rankings, we can identify two stakeholders, the people represent the rankers, and the people represent the rankees. The people represent the rankers, such as government officers behind the criminal justice systems~\cite{COMPAS}, or the hiring manager controls the recruiting system~\cite{hiregenderbias} interact directly with the rankees. The people represent the rankees such as university administrators, company board members who intent to advance in the ranking and seeking reputation and financial gains. The goals between rankers' and rankees' representatives may differ and affect the ranking results by involving in different part of the ranking process.
\par \noindent \textbf{Are data subjects empowered?} 
Government officers behind the criminal justice systems~\cite{COMPAS}, or the hiring manager who controls the recruiting system~\cite{hiregenderbias} have disproportionate power over people as rankees or data subjects. In the case of institutional rankings, universities can at least make educated guesses about what data to report that affect their rankings. 
For stakeholders such as university ranking publishers, it is a typical process to question the quality of the data reported by the university administrators. When the data reported from the university does not satisfy the requirements or the university does not comply with the requirements, the particular university or rankee may be omitted from the ranking. However, such interaction in the data collection stage is not transparent~\cite{castillo2019fairness} between the rankers and rankees.
\par \noindent \textbf{Who has agency?}
% Not only is the data collecting stage opaque, but the ranker designing stage also lacks explanation. For the example of university rankings, 
Ranking publishers typically do not provide the closed-form mathematical formulas but instead a methodology
% description As the majority of the university rankings are score-based~\cite{ammar2011ranking}, the methodology may 
describing the factors considered in the ranking formulation such as \textit{Graduate and Retention rates} 22\% or \textit{Faculty Resources} 20\%~\cite{USnews}. 
Although they are the decision-makers, students and their parents may not have agency in the ranking process, but rather are influenced by the social impact of the popular rankings.
The rankers and decision-makers may represent the same stakeholders in a different use case. For example, hiring managers may design the ranking formula for job candidates. However, the rankees' representatives, potential or current employees, do not have the agency or the knowledge for evaluating the rankers, and are often prone to biases towards candidates~\cite{hiregenderbias, blackstudents}.
% . The issues aroused recently in the media about the gender bias~\cite{hiregenderbias}, and racial discrimination~\cite{blackstudents} in such rankings may become more severe if the involvement from the rankees' side remains insufficient.
\par \noindent \textbf{Can the decision-makers be held accountable?}
We might make an assumption that hiring managers are equipped with the knowledge to understand and evaluate the behavior of rankers. However, such an assumption is questionable. Note that some rankers learn the ranking of items, while others are simpler, based on a scoring formula that computes a weighted sum of attribute values of the items being ranked.  Yet, even with a simple scoring formula, the intention of the hiring manager may not be followed, or it may even be reversed in certain subsets of the data ~\cite{stoyanovich2020imperative}. The hiring managers may evaluate the ranker based on accuracy in the top candidates and gain false confidence in the ranker's accuracy in the lower ranked candidates, raising accountability concerns. The situation worsens when the ranker uses a complex learned model, as is becoming increasingly widespread in the industry.

\begin{tcolorbox}
\textbf{Statement 2:} Technological innovation and acceptance need to be guided by democratizing interpretability approaches where ranking processes are made explainable to people who are impacted and, as a result, the outcomes adhere to higher ethical standards and new legislations~\cite{act22}.
\end{tcolorbox}
% \aritra{the section name should abstract a particular position statement we make. This is just a sample}

% The involvement gap in the ranking process is severe, compounded by the lack of carefully designed and implemented ranking explanations.
% As rankings become ubiquitous in the social-technical world, the lack of explanation of rankings is unnoticed or even normalized. 
% \par \noindent \textbf{No explanation.}
% Modern internet users interact with webpage rankings from search engines, restaurant rankings from location-based commendation apps, even the most "attractive" individuals from dating apps. 
% However, as the trust~\cite{schillo2000using} gradually increase between decision-makers and the rankers, there is usually no explanation of the underlying rankers. 
% The companies or organizations of such rankers may have no intention to explain the underlying ranking mechanism due to privacy or proprietary reasons. However, \cite{stoyanovich2020imperative} points out that it may lead to certain groups, usually benefiting from the ranking, over-trusting the ranking, or the disadvantaged groups, refusing to trust the ranking and sacrifice the benefit of rankings. 
\par \noindent \textbf{Are institutional explanations sufficient?}
Some representatives of the rankers, such as the university ranking publishers, may provide explanations via disclosing the methodology of their ranking process. However, the methodology is typically lengthy yet lacks details. There is neither a way to validate the methodology nor to modify the methodology from rankees' or decision-makers' perspective, who may not completely agree with the methodology's emphasis. For example, a student may want to put more emphasis on \textit{teaching} than \textit{research} when choosing universities. 
% Decision-makers may identify some measures and metrics used to produce rankings in methodology. They may access those data from the government's open data portals. But usually, the rankings may involve subjective surveys conducted by the representative of the rankers. 
\par \noindent \textbf{Can we explain inaccessible rankers using available data?}
Publicly available data~(e.g., about hiring practices or college admissions) might provides opportunities to get data-driven explanations about decisions made by inaccessible rankers. For example, decision-makers may build proxy ranking models using machine learning~\cite{liu2009learning} approaches to mimic the behavior of rankers. Decision-makers can learn about the unknown rankers by manipulating the inputs manually and observing the output, as shown in Google's What-if tool~\cite{wexler2019if}. Alternatively, such manipulating of the input data can be conducted systematically with proper statistical sampling approaches to derive each data attribute's impact on advancing a rankee's rank position. The recent development of model-agnostic instance-wise explanation methods such as LIME~\cite{ribeiro2016should} and SHAP~\cite{lundberg2017unified} is proven to be effective in explaining behaviors of complicated Machine Learning/Deep Learning models without knowing the model process under the hood. Lack of effective explanation methods leads to concerns of fairness, accountability and transparency~\cite{DBLP:journals/pvldb/AsudehJMS18,DBLP:conf/sigmod/GuanAMJSM019,DBLP:conf/sigmod/YangSAHJM18}. 
% With the increasing trend of using ML and DL models in the industry, the rankers' representatives may also benefit from using a data-driven explanation approach to evaluate their rankers. The rankers' representatives, who have access to the entire data, may further evaluate their rankers on the actual datasets and scrutiny the rankers under real data, in different rank neighborhoods, and the emerging ethical issues in AI such as fairness, accountability, and transparency~\cite{NYCADS}. 
\par \noindent \textbf{How can we abstract information that enlighten people about rankers?}
% Different approaches of explanation depending on the availability of the ranking input data, the complexity of the rankers, and the ranking results. 
Even though the data-driven explanation approaches are promising, we need information abstractions that are informed by data science, social science, and principles of human-data interaction.
Nutritional label for ranking~\cite{DBLP:conf/sigmod/YangSAHJM18} is a promising approach in that direction, where explanations about score-based rankings are communicated at different levels of abstraction, inspired from food science, where nutrition facts, ingredients, and recipes, communicate different levels of detail about food, without requiring expert knowledge from the consumer.

\section*{Acknowledgements}
\label{sec:ack}
This work was supported in part by NSF Grants No. 1916505 and 1934464.

\bibliographystyle{ACM-Reference-Format}
\bibliography{REFERENCE}

\end{document}